\documentclass{article}

\usepackage{spconf}

\usepackage{
    amsfonts,
    amsmath,
    color,
    xcolor,
    colortbl,
    commath,
    graphicx,
    hyperref,
    multirow,
    soul,
    wasysym
}

\definecolor{tablecolor}{cmyk}{0,0,0,0.12}

\title{High-Fidelity Neural Phonetic Posteriorgrams}
\name{Cameron Churchwell$^*$, Max Morrison$^*$, Bryan Pardo} 
\address{Northwestern University, Evanston, IL, USA}

\begin{document}
\ninept
\maketitle
\def\thefootnote\footnotetext{Equal contribution}\def\thefootnote{\arabic{footnote}}


\begin{abstract}
A phonetic posteriorgram (PPG) is a time-varying categorical distribution over acoustic units of speech (e.g., phonemes). PPGs are a popular representation in speech generation due to their ability to disentangle pronunciation features from speaker identity, allowing accurate reconstruction of pronunciation (e.g., voice conversion) and coarse-grained pronunciation editing (e.g., foreign accent conversion). In this paper, we demonstrably improve the quality of PPGs to produce a state-of-the-art interpretable PPG representation. We train an off-the-shelf speech synthesizer using our PPG representation and show that high-quality PPGs yield independent control over pitch and pronunciation. We further demonstrate novel uses of PPGs, such as an acoustic pronunciation distance and fine-grained pronunciation control.
\end{abstract}




\section{Introduction}

\begin{figure*}[ht]
    \centering
    \includegraphics[width=\linewidth]{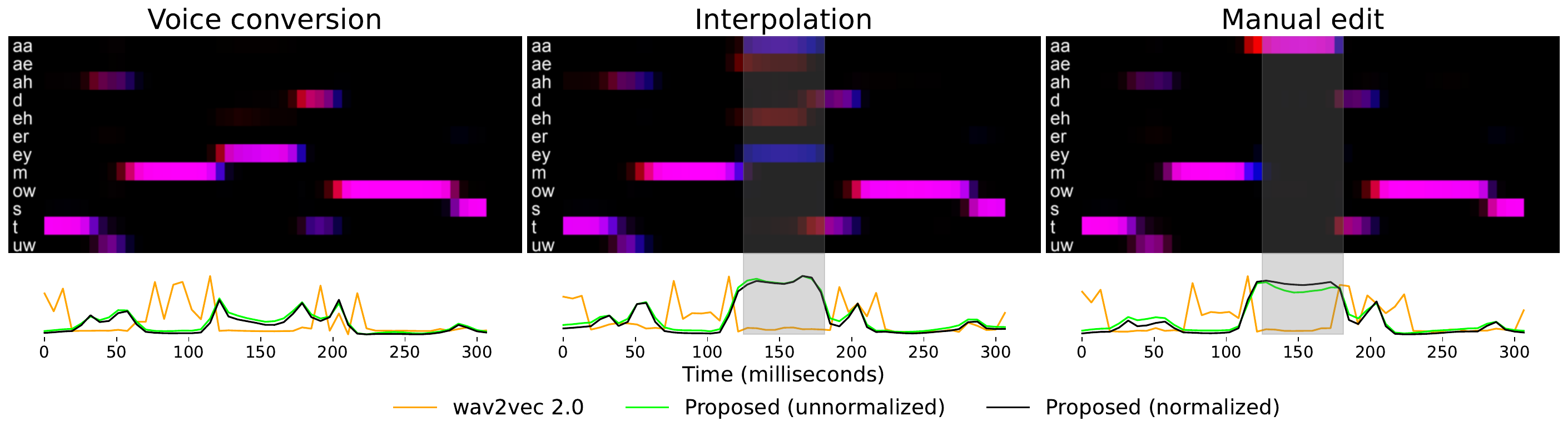}
    \vspace{-2.3em}
    \caption{\textbf{Pronunciation interpolation and distance $|$} We train a VITS~\cite{vits} speech synthesizer on our interpretable PPGs and use it for \textbf{(left)} voice conversion, \textbf{(center)} pronunciation interpolation, and \textbf{(right)} manual phoneme editing. \textbf{(top)} We visualize overlapping PPGs of a recording of the word "tomato" (\textcolor{blue}{blue}) and inferred from the synthesized speech (\textcolor{red}{red}). For readability, phoneme rows in the PPGs with maximum probability $<10\%$ are omitted. The accurate reconstruction of PPGs (\textcolor{magenta}{magenta}) indicates preservation of (potentially edited) phonetic content in the generated speech. In the center, the input (\textcolor{blue}{blue}) PPG is interpolated halfway between the left and right PPGs using SLERP~\cite{shoemake1985animating}.  Note that the reconstruction of interpolating ``ey'' (\textbf{left}) and ``aa'' (\textbf{right}) is ``ae'' or ``eh'' (\textbf{center}). This is consistent with interpolating vowels in formant space (F1, F2 - F1)~\cite{ladefoged2014course} and indicates that one pronunciation can be represented more than one way in a PPG. \textbf{(bottom)} Pronunciation distances between synthesized speech and the original audio. Our proposed distance (Section~\ref{sec:distance}) is more robust to resynthesis artifacts and accurately captures pronunciation interpolation without a transcript.}
    \label{fig:interpolation}
    \vspace{-1em}
\end{figure*}

The phonetic posteriorgram (PPG) ~\cite{originalppg} is a time-varying categorical distribution over acoustic units of speech (e.g., phonemes). PPGs have enabled  voice conversion without changing pronunciation \cite{voiceconversion, 
liu2021any, kovela2023any} (Figure \ref{fig:interpolation}, left). As such, all five top entries to the 2020 Voice Conversion Challenge~\cite{zhao2020voice} utilize PPGs. Beyond voice conversion, text-to-speech (TTS) systems that predict PPGs from text as an intermediate have shown improved pronunciation relative to predicting speech directly from text~\cite{adavits} and have enabled accent conversion~\cite{foreignaccentconversion}. 

Speech synthesis tasks (e.g., text-to-speech~\cite{wang2023neural}) typically use as an input representation the sequence of phonemes indices, extracted from the transcript via a grapheme-to-phoneme process. This representation does not specify the exact pronunciation of each phoneme, or its duration; however, phoneme durations can be inferred from ground truth speech and corresponding transcripts (e.g., via forced phoneme alignment~\cite{charsiu}). In contrast, PPGs accurately represent pronunciation, preserve alignment, and permit training of speech synthesizers without access to the speech transcript---requiring a transcript only for the initial training of the generalizable PPG model. 

Diphone and triphone models can be used to represent transitions between phonemes in an interpretable representation~\cite{diphone}, but are not designed to represent ambiguity when pronunciation falls at the border between similar phonemic categories. As with phoneme indices, training and generation using diphones or triphones requires either access to the speech transcript---in which case the pronunciation and phoneme durations are inferred and contain inaccuracies---or manual diphone or triphone annotations produced by an expert.

Prior works have noted that pronunciation is preserved during voice and accent conversion when using representations like intermediate activations of ASR systems~\cite{liu2021any} or distributions over learned latent variables~\cite{foreignaccentconversion}---and have even used the term PPG to refer to some of these representations. While all of these are multi-dimensional, continuous-valued representations, none of these representations permit the interpretability and control afforded by true PPGs built upon interpretable phonetic categories.

No prior work has closely evaluated the impact of input representations for PPGs on downstream speech fidelity or the entanglement of pronunciation and prosody. No prior work has demonstrated fine-grained user control of pronunciation, such as interpolation between phonemes. This is useful for correcting mispronunciations or accents within podcasts, video games, and film dialogue as well as measuring acoustic pronunciation distance for, e.g., evaluating voice conversion and speech editing. Our contributions are as follows:

\begin{itemize}
    \item \textbf{(Contribution 1)} We propose an interpretable PPG representation (Section~\ref{sec:model}) that exhibits competitive pitch modification accuracy relative to existing, non-interpretable speech representations (Section~\ref{sec:disentangle})\footnote{Audio examples:
    \texttt{\href{https://maxrmorrison.com/sites/ppgs}{maxrmorrison.com/sites/ppgs}}
    }.
    \item \textbf{(Contribution 2)} We propose an interpretable speech pronunciation distance (Figure~\ref{fig:interpolation}; bottom) based on the Jensen-Shannon divergence between PPGs. This is a time-aligned, language-agnostic alternative to word error rate (Section~\ref{sec:distance}).
    \item \textbf{(Contribution 3)} We are the first to demonstrate that interpretable PPGs enable fine-grained pronunciation control, including interpolation (Figure~\ref{fig:interpolation}; top), regex-based accent conversion, and automatic onomatopoeia (Section~\ref{sec:control}).
\end{itemize}

\noindent
To facilitate future research, we release our code\footnote{Code:
\texttt{\href{https://github.com/interactiveaudiolab/ppgs}{github.com/interactiveaudiolab/ppgs}}}
and speech representations as \texttt{ppgs}, an MIT-licensed, pip-installable Python module for training, evaluating, and performing inference with PPGs.


\section{Network architecture}
\label{sec:model}


Neural networks for inferring PPGs take a sequence of audio features (see Section \ref{sec:audio_input_represenatations}) at some frame resolution (e.g., ten milliseconds) and produce a categorical distribution over phonemes at each frame. Prior work has not thoroughly investigated what input representation maximizes PPG performance. We address this by selecting a representative, high-performing network architecture and, for each of a variety of audio input encodings (Section~\ref{sec:audio_input_represenatations}), train our selected network architecture to produce PPGs. We compare the resulting PPGs with each other, as well as other recent speech representations.

Our network architecture consists of an input convolution layer, five Transformer encoder layers (self-attention and a feed-forward network)~\cite{attention}, and an output convolution layer that produces a categorical distribution via softmax activation over 40 phonemes (including silence) from the CMU Pronunciation Dictionary phoneme set~\footnote{\texttt{\href{http://www.speech.cs.cmu.edu/cgi-bin/cmudict}{speech.cs.cmu.edu/cgi-bin/cmudict}}}. We use a kernel size of five for the input and output convolution layers. Our Transformer layers use two attention heads and 512 channels. For each representation, we selected the number of layers and channels via hyperparameter search on a heldout validation partition from Common Voice~\cite{commonvoice} (Section~\ref{sec:data}). We fixed the number of channels at 128; trained using 3, 4, 5, and 6 layers; and then fixed the number of layers at best of these values and trained models with 128, 256, 512, and 1024 channels, selecting the best of these. In the event of divergence from overparameterization (i.e., \textit{gradient confusion}~\cite{sankararaman2020impact}), we allow one reload from checkpoint. We find 5 layers and either 256 (for EnCodec and Mel spectrograms) or 512 (for all others) channels to be optimal. We use an Adam optimizer~\cite{kingma2015adam} with a learning rate of $2e^{-4}$ to optimize categorical cross entropy loss between predicted and ground truth phoneme categories at each ten millisecond frame. We train for 200,000 steps using a variable batch size~\cite{gonzalez2023batching} of up to 150,000 frames per batch. 

We synthesize speech by training VITS~\cite{vits} on each representation with and without converting to PPGs. We replace the upsampled phoneme features with our representation and concatenate with pitch inferred with FCNF0++~\cite{morrison2023cross}, clipped to 50-550 Hz, evenly quantized in base-two log-Hz to 256 bins, and embedded in a 64-dimensional embedding table.


\section{Evaluation}
\label{sec:evaluation}
 
We design our evaluation to answer three questions: (1) What audio input representation is best for producing accurate PPGs? (Sections~\ref{sec:audio_input_represenatations}, \ref{sec:accuracy}), (2) How good are PPGs at disentangling pitch and pronunciation? (Section~\ref{sec:disentangle}), and (3) Are our proposed PPGs suitable for high-quality speech synthesis? (Section~\ref{sec:subjective}). We further perform correlation analysis between framewise accuracy and subjective preference to establish an objective evaluation proxy for costly subjective evaluation (Figure~\ref{fig:mushra}).

\subsection{Audio input representations}
\label{sec:audio_input_represenatations}

Representations are computed at a hopsize of ten milliseconds (ms) and a sample rate of 16,000 Hz, unless otherwise stated. Baseline neural representations are pretrained using original implementations.


\noindent
\textbf{Mel spectrogram} [80 channels] $|$ Spectrograms are a common representation for speech research tasks. We use log-energy magnitude spectrograms computed from the raw audio with a window size of 1024 and bin the frequency channels into 80 Mel-spaced bands.

\noindent
\textbf{Wav2vec 2.0}~\cite{wav2vec2} [768 channels] $|$ Wav2vec 2.0 is a neural speech encoder that achieves state-of-the-art ASR Phoneme Error Rate (PER) when fine-tuned on TIMIT. Wav2vec 2.0 uses a 20 ms hopsize. We apply nearest neighbors interpolation (which outperforms linear) to double the number of timesteps to a 10 ms hopsize.

\noindent
\textbf{Charsiu}~\cite{charsiu} [768 channels] $|$ Charsiu appends a convolutional layer to a pretrained wav2vec 2.0 base model that upsamples from the 20 ms hopsize to a 10 ms hopsize. The wav2vec 2.0 feature encoder is frozen and the rest of the model is fine-tuned to maximize a categorical cross entropy loss over ground truth derived via grapheme-to-phoneme and forced alignment~\cite{mcauliffe2017montreal}. We use the \texttt{W2V2-FC-10ms} model, which achieves state-of-the-art in forced alignment~\cite{charsiu}.

\noindent
\textbf{ASR bottleneck}~\cite{liu2021any} [144 channels] $|$ This is a pretrained ASR model with an encoder-decoder architecture. We use the bottleneck features output by the pretrained encoder, which is also used in voice conversion and TTS for its pronunciation-preserving qualities~\cite{kovela2023any, adavits}.

\noindent
\textbf{EnCodec}~\cite{defossez2022high} [128 channels] $|$ EnCodec converts audio into codebook indices of 32 codebooks---each containing 1024 codes and 128 channels---and then performs an element-wise sum over codebooks. EnCodec achieves competitive results on low-dimensional, invertible speech representation learning~\cite{defossez2022high} and text-to-speech~\cite{wang2023neural}.


\subsection{Data}
\label{sec:data}
We train on Common Voice 6.1~\cite{commonvoice} and perform objective evaluation of phoneme accuracy using a held-out partitions of Common Voice, as well as the full CMU Arctic~\cite{arctic} and TIMIT~\cite{timit} datasets. We use open-source Common Voice alignments produced by Charsiu~\cite{charsiu}. The transcripts for Arctic and TIMIT are phonetically balanced and manually time-aligned. We partition Common Voice into train/valid/test partitions of proportions 80\%/10\%/10\%. We train and evaluate our VITS~\cite{vits} speech synthesizers on VCTK~\cite{vctk}.



\begin{figure}[t]
    \centering
    \includegraphics[width=\linewidth]{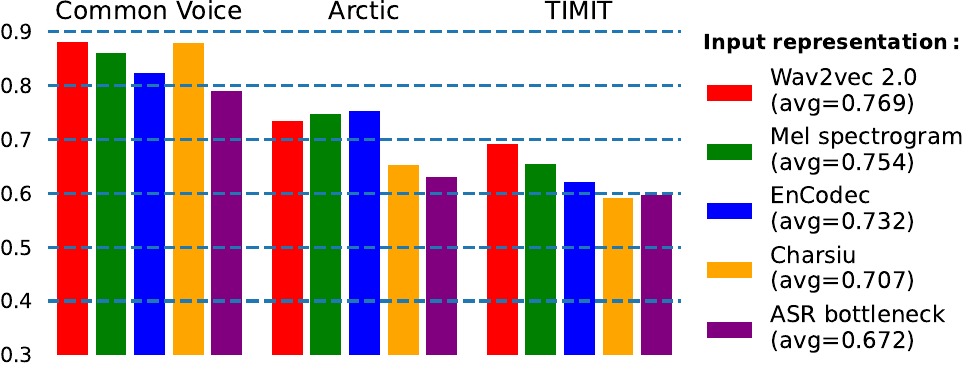}
    \vspace{-2.1em}
    \caption{\textbf{Average framewise phoneme accuracy $|$} Accuracy of PPGs computed from five input representations. The wav2vec 2.0~\cite{wav2vec2} input representation has the best PPG accuracy when averaged over all datasets (see legend). \textbf{N.B.,} The base wav2vec 2.0 model of Charsiu~\cite{charsiu} was trained on some of our Common Voice test partition as well as the TIMIT training partition, making Charsiu's results on those datasets unreliable upper bounds.}
    \label{fig:accuracy}
    \vspace{-1em}
\end{figure}


\subsection{Objective evaluation of phoneme accuracy}
\label{sec:accuracy}

While PPGs allow a more nuanced representation than discrete phonemes, any network that infers PPGs from audio should broadly agree with high-quality, aligned phonetic transcriptions. We perform objective evaluation to determine the extent to which our PPG representations computed from each input representation (Section~\ref{sec:audio_input_represenatations}) accurately predict ground truth phoneme categories. We evaluate the framewise phoneme accuracy, or the proportion of frames where the ground-truth phoneme is assigned highest probability by the model. We perform evaluation on our test partition of Common Voice, as well as all of Arctic and TIMIT. These framewise phoneme accuracies are not directly comparable to unaligned connectionst temporal classification (CTC) phoneme error rates (PER) for ASR~\cite{wav2vec2}, nor framewise accuracies of models trained on heldout speakers and datasets~\cite{graves2005framewise}. Results can be found in Figure~\ref{fig:accuracy}.


\subsection{Objective evaluation of disentanglement}
\label{sec:disentangle}

We evaluate disentanglement of pitch and pronunciation by demonstrating pronunciation invariance when pitch-shifting by $\pm 200$ cents. Given pitch values $y$ and $\hat{y}$ in Hz, cents is the perceptually linear ratio $\abs{1200 \log_2 (y / \hat{y})}$; one musical semitone is 100 cents. We use 100 utterances (from 10 speakers; 5 male and 5 female) in the VCTK~\cite{vctk} dataset and encode each into 10 speech representations: 5 input representations (Section~\ref{sec:audio_input_represenatations}) and 5 corresponding PPG representations inferred from input representations. We train 10 VITS~\cite{vits} models---one for each representation---and use each model to perform synthesis using the 100 selected utterances. We use three error metrics: (1) pitch error, (2) word error rate (WER), and (3) our proposed  PPG-based pronunciation distance ($\Delta \text{PPG}$) described in Section~\ref{sec:distance}.

We measure pitch error as the average framewise error in cents: $\Delta\cent(y, \hat{y}) = \frac{1200}{|\mathcal{V}|} \sum_{t \in \mathcal{V}} \abs{ \log_2 (y_t / \hat{y}_t)}$, where $y = y_1, \dots, y_T$ is the ground truth frame resolution pitch contour in Hz; $\hat{y} = \hat{y}_1, \dots, \hat{y}_T$ is the predicted pitch contour in Hz; and $\mathcal{V}$ are the time frames where both the original and re-synthesized speech contain a pitch (i.e., when the entropy-based periodicity exceeds 0.1625~\cite{morrison2023cross}). We measure WER as a fraction between zero and one by using Whisper-V3~\cite{radford2023robust} to transcribe the generated speech and comparing to ground truth transcripts.

Results of this objective evaluation are in Table~\ref{tab:disentangle}. We see Mel spectrograms and EnCodec fail to pitch-shift due to entanglement, producing intelligible pronunciation at the original pitch. Wav2vec 2.0~\cite{wav2vec2} and the Charsiu forced aligner~\cite{charsiu} both produce state-of-the-art disentanglement---outperforming the widely-used ASR bottleneck~\cite{liu2021any} in pronunciation accuracy. PPGs computed from Mel spectrograms and EnCodec~\cite{defossez2022high} outperform wav2vec 2.0 and Charsiu in pitch disentanglement, but with less accurate pronunciation. Addressing this pronunciation error gap is an important research direction, as Charsiu, wav2vec 2.0, and ASR bottleneck are non-interpretable, use at least an order of magnitude more channels per frame, and do not enable the properties discussed in Section~\ref{sec:properties}.

\begin{table}[t]
\centering
\begin{tabular}{lrcc}
& \textbf{$\boldsymbol{\Delta\cent}\downarrow$} & \textbf{WER$\downarrow$} & \textbf{$\boldsymbol{\Delta \text{PPG}}\downarrow$} \\
\rowcolor{tablecolor}
\textbf{Mel spectrogram} & 207.7  & 0.0239 & 0.1063 \\
\quad \textbf{PPG} & 56.0 & 0.0744 & 0.2014 \\
\rowcolor{tablecolor}
\textbf{Wav2vec 2.0~\cite{wav2vec2}} & 57.2 & 0.0244 & 0.1528\\
\quad \textbf{PPG} & 59.5 & 0.0910 & 0.2616 \\
\rowcolor{tablecolor}
\textbf{Charsiu~\cite{charsiu}} & 59.2 & 0.0214 & 0.1652 \\
\quad \textbf{PPG} & 61.8 & 0.5074 & 0.5245 \\
\rowcolor{tablecolor}
\textbf{ASR bottleneck~\cite{liu2021any}} & 55.8 & 0.0558 & 0.2026 \\
\quad \textbf{PPG} & 65.9 & 0.2779 & 0.4164 \\
\rowcolor{tablecolor}
\textbf{EnCodec~\cite{defossez2022high}} & 183.8 & 0.0260 & 0.1654 \\
\quad \textbf{PPG} & 56.5 & 0.1018 & 0.2014 \\
\end{tabular}
\vspace{-.2em}
\caption{\textbf{Pitch and pronunciation disentanglement $|$} Results are averages over pitch-shifting down ($\boldsymbol{-200\cent}$) and up ($\boldsymbol{+200\cent}$) using VITS~\cite{vits} with either one of our input features or proposed PPG representations computed from each input representation.}
\label{tab:disentangle}
\vspace{-1.1em}
\end{table}


\subsection{Subjective evaluation of speech synthesis quality}
\label{sec:subjective}

\begin{figure}[t]
    \centering
    \includegraphics[width=\linewidth]{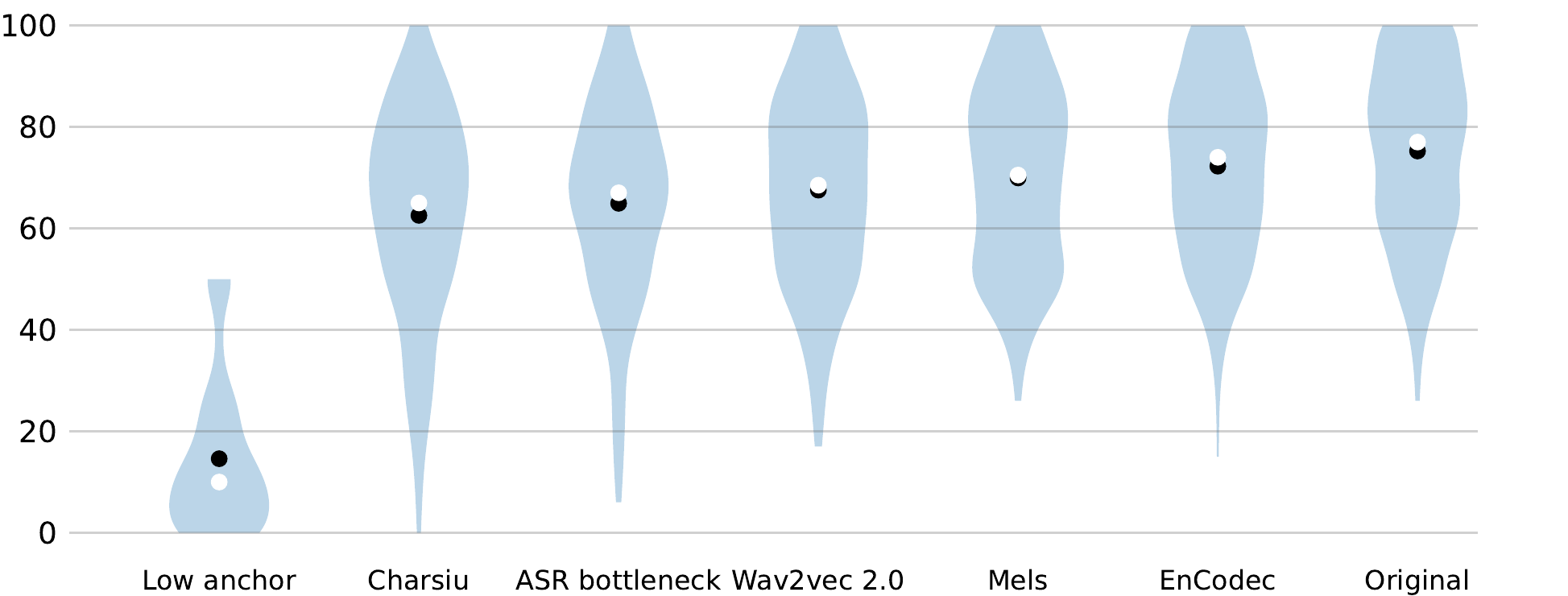}
    \caption{\textbf{Crowdsourced subjective evaluation results $|$} \textbf{(top)} Reconstruction quality of speech synthesized from PPGs inferred from five input representations, as well as high- and low-anchors. White dots are medians and black dots are means. A Wilcoxon signed-rank test gives $p=0.02$ between original speech and speech reconstructed using PPGs inferred from EnCodec. Interpretable PPGs inferred from EnCodec significantly outperform $(p<0.05)$ PPGs inferred from all other representations except Mel spectrograms ($p=0.25$).}
    \label{fig:mushra}
    \vspace{-1em}
\end{figure}

To validate that PPGs allow high-fidelity speech generation, we use Reproducible Subjective Evaluation (ReSEval)~\cite{morrison2023reproducible} to perform a subjective listening test in the Multiple Stimuli with Hidden Reference and Anchor (MUSHRA)~\cite{mushra} format. Each participant performs 10 MUSHRA trials. In each trial, one of the 100 utterances used in the objective evaluation (Section~\ref{sec:disentangle}) is reconstructed (using the original pitch contour without pitch-shifting) from PPGs computed from each of the five input representations. The reconstructions are rated in comparison to each other on a 0-100 quality scale using a set of sliders. Two references are also included in the comparison set: (1) the high-quality original speech audio (the high anchor) and (2) a low-quality, 4-bit quantization of the original audio (the low anchor). We recruited 50 participants on Amazon Mechanical Turk, filtering for US residents with an approval rating of at least 99\% and at least 1,000 approved tasks. We paid annotators \$3.50 for an estimated 15 minutes of work (\$14 per hour). We filtered out 26 annotators that either failed the listening test or rated the low anchor (4-bit quantized audio) as higher quality than the high anchor (original audio). Results are in Figure~\ref{fig:mushra}.


\section{Properties of Phonetic Posteriorgrams}
\label{sec:properties}

We discuss additional evidence gathered in this work supporting the existence of useful and interesting properties of our interpretable PPG features.

\subsection{PPGs encode acoustic pronunciation distance}
\label{sec:distance}

\begin{figure}[t]
    \centering
    \includegraphics[width=\linewidth]{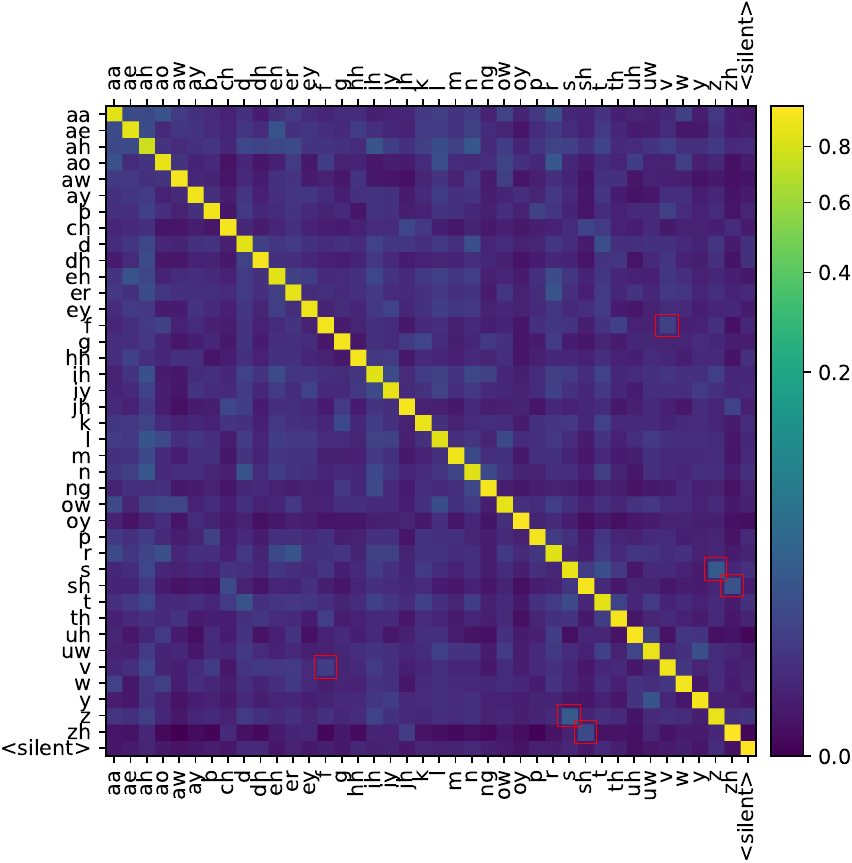}
    \caption{\textbf{Acoustic phoneme similarities $|$} Row $x$ column $y$ is $\mathcal{S}_{x, y} = \mathbb{E}\left[\lambda_y G_{y, t}; \lambda_x G_{x, t} \geq \lambda_z G_{z, t} \, \forall z \right]$, the average class-weighted probability assigned to phoneme $y$ when phoneme $x$ is the maximum model prediction. Averages are taken over all frames of our validation partition of Common Voice~\cite{commonvoice} using our PPG model trained with class-balancing on Mel spectrogram inputs. \textcolor{red}{Red} boxes show that the corresponding unvoiced fricative (/f/, /s/, /sh/) to each voiced fricative (/v/, /z/, /zh/) is assigned relatively high probability, and vice versa. Class-balanced training and class-weighting are used to remove column banding indicative of natural phoneme frequency.}
    \label{fig:confusion}
    \vspace{-1.5em}
\end{figure}

We propose an interpretable distance measure of framewise pronunciation error. Let $G \in \mathbb{R}^{|P| \times T}$ be a phonetic posteriorgram on phoneme set $P$ and time frames $T$, such that $G_{p, t}$ is the inferred probability that the speech in frame $t$ is phoneme $p$. By default, our PPG training is not class-balanced, and some phonemes are significantly more likely to occur in the dataset (e.g., ``aa'' occurs far more often than ``zh''). To prevent this from imposing bias on our proposed distance, we train a class-balanced PPG model using class weights $\lambda_i=\min_j F_j / F_i$ to weight the relative contribution of each phoneme to the training loss, where $F_x$ is the number of frames where phoneme $x$ is ground truth. We extract from this class-balanced model an interpretable representation of similarity $\mathcal{S} \in \mathbb{R}^{|P| \times |P|}$ between phonemes (Figure ~\ref{fig:confusion}). Our proposed acoustic pronunciation distance $\Delta \text{PPG}$ can be stated as follows.
\begin{equation}
    \Delta \text{PPG}(G_t, \hat{G}_t) = \text{JS}(\mathcal{S}^\gamma G_t, \mathcal{S}^\gamma \hat{G}_t)
\end{equation}
$\text{JS}$ is the Jensen-Shannon divergence. We tune $\gamma$ on our validation partition to maximize the Pearson correlation with WER. Using the test data from Table~\ref{tab:disentangle} and optimal hyperparameter $\gamma=1.20$, $\Delta \text{PPG}$ demonstrates strong Pearson correlation with WER ($r=0.697$; $n=2000$; $p=1.76\times 10^{-291}$).

We further inspect the behavior of our proposed phoneme distance to capture frame-level pronunciation differences during pronunciation editing. We use as a baseline the dynamic time warping (DTW) between wav2vec 2.0 latents, which has been shown to outperform spectral-based and transcript-based speech variation distances~\cite{bartelds2022neural}. Our audio is already aligned, so we replace DTW with framewise L2 distance. Figure~\ref{fig:interpolation} (bottom) demonstrates the behavior of each pronunciation distance during voice conversion (left), pronunciation interpolation (center), and manual pronunciation editing (right). While wav2vec 2.0~\cite{wav2vec2} enables disentanglement (Table~\ref{tab:disentangle}), it fails to detect aligned pronunciation differences captured by our proposed, interpretable pronunciation distance based on the JS-divergence between PPGs.

\subsection{PPGs enable fine-grained pronunciation control}
\label{sec:control}

While prior works have demonstrated that PPGs enable conversion between accents~\cite{foreignaccentconversion}, no prior work has demonstrated interpretable, fine-grained user control of speech pronunciation. We present the first such example by demonstrating interpolation between two common pronunciations of a single word within an utterance (Figure~\ref{fig:interpolation}; top). We use spherical linear interpolation (SLERP)~\cite{shoemake1985animating} for interpolating PPGs to maintain a valid distribution. As described in Section~\ref{sec:distance}, we use as pronunciation reconstruction error the JS divergence between the input, interpolated PPG and the corresponding PPG inferred from the generated audio (Figure~\ref{fig:interpolation}; top). When trained to synthesize speech from pitch and interpretable PPGs, models such as VITS~\cite{vits} acquire a diverse set of affordances for speech editing, including existing controls (voice conversion, pitch-shifting, and singing voice transfer) as well as novel fine-grained pronunciation control. To further demonstrate the novel pronunciation control enabled by interpretable PPGs, we propose two novel types of speech editing: (1) interpretable accent conversion via regex-based editing of monophone, diphone, and triphone sequences contained in the PPG and (2) automatic onomatopoeia, in which speech is synthesized to mimic non-speech audio in a target voice. Audio examples of all of these speech editing controls are on 
\href{https://www.maxrmorrison.com/sites/ppgs/}{our companion website}.


\section{Conclusion}
\label{sec:conclusion}

Phonetic posteriorgrams (PPGs) are time-varying distributions over phoneme categories that capture fine-grained pronunciation information. In this work, we propose an interpretable PPG representation with competitive pitch disentanglement relative to widely-used, non-interpretable representations \textbf{(Contribution 1)}. We discover novel properties of interpretable PPGs, such as an acoustic phoneme distance \textbf{(Contribution 2)} and fine-grained pronunciation control \textbf{(Contribution 3)}. Future work may explore the manifold of valid interpolations and evaluate our PPGs in downstream tasks such as accent coaching and mispronunciation detection.


\bibliographystyle{IEEEbib}
\bibliography{refs}


\end{document}